# The benefits of structural disorder in natural cellular solids

D. Aranguren van Egmond[1,*,†], B. Yu[2], S. Choukir[3], S. Fu[1], C.V. Singh[1,3], G.D. Hibbard[1,*], B.D. Hatton[1,3*]

[1]Department of Materials Science and Engineering, University of Toronto, Toronto, Ontario, Canada M5S 3E4.

[2]Department of Materials Science and Engineering, McMaster University, Hamilton, Ontario, Canada L8S 4L7.

[3]Department of Mechanical and Industrial Engineering, University of Toronto, Toronto, Ontario, Canada M5S 3G8.

*Correspondence to: derek.aranguren@alum.utoronto.ca; benjamin.hatton@utoronto.ca;

†Present address: National Research Council of Canada, 100 Sussex Drive, Ottawa, Ontario, Canada K1A 0R6

**Abstract:** Architected cellular materials in nature such as trabecular bone and woods exhibit disordered porous structure that is neither fully random, nor perfectly ordered, and has not been systematically quantified before. We have used Voronoi cell patterning to quantify disorder for a range of biological and engineered cellular materials, using a 'disorder parameter' ($\delta$). Ranges of disorder appear to be typical to certain plant, animal and fungi cellular materials. Using 3D printing and numerical methods, we demonstrate experimentally there is a range of 'pseudo-order' ($\delta=0.6 – 0.8$) which exhibits a >30% increase in fracture toughness (and equivalent strength) compared to fully-ordered, hexagonal honeycombs ($\delta=1.0$) of equal density. Our results suggest 'tailored disorder' as a new design paradigm for architected materials to improve damage tolerance, and an evolutionary advantage for the pseudo-order of biological cellular materials.

Porous, biological cellular materials such as wood (plant ground tissue), trabecular bone, corals, diatoms, and dentin combine complex biological functions with structural roles, such as skeletal support and impact protection (*1, 2*). Biological cellular materials typically feature highly complex structural hierarchies, from nano- to macroscale that enable optimization of both strength and toughness (flaw tolerance) simultaneously (*3-9*). The hierarchy of bone, for example, ranges over 9-10 orders of magnitude in length scale, from the molecular level to the macroscale (*10, 11*). Recently, there is great interest in the optimization of internally 'architected' engineering materials, featuring design across several length scales. Architected materials, be they biological or engineered, feature mechanical properties defined by composition and multiscale hierarchical geometry (*3*).

As illustrated in a cross-section view of a whale vertebra (trabecular bone) (**Fig 1a**), biological cellular materials are generally disordered in the spatial distribution of cells (pores) and resulting cell-size variations. This disorder contrasts with engineered, structural cellular solids, such as hexagonal honeycombs and micro-truss materials, which are typically highly ordered. Since nature is clearly capable of complex material design, we suggest that highly ordered structure should be readily encountered in biological cellular materials if that order was advantageous. Instead, we hypothesize that structural disorder in cellular solids may itself provide certain benefits in mechanical performance, principally as a toughening mechanism.

Certainly, biological materials display a rich domain of design strategies for optimizing mechanical performance, e.g.: 1) the layering of bending-dominated and stretching-dominated microstructures, 2) weak interfaces, 3) functionally gradient materials, and 4) crack bridging by fibrous elements (*7, 10-17*). For example, to limit catastrophic failure the hierarchical organic/inorganic structures within bone and nacre significantly increase the flaw tolerance and effective work of fracture compared to engineered ceramics (*13-15, 18, 19*).

Surprisingly, the spatial disorder of biological cellular materials and its role in mechanical performance have not been analyzed or quantified previously. Herein, we have collected cross-sectional images (our own specimens and sourced from literature, **Table S1**) of biological cellular materials across a wide range of species and length scales (**Fig 1a-f**). We use the Voronoi tessellation as a model to both simulate and quantify disorder, as cell size variation (*20*). Through an image analysis algorithm, we generated representative Voronoi structures for each image using the centroids of the open cells as nucleation sites (details available in Supplementary Materials, SM). The regular hexagonal honeycomb (RHH) is considered as a special case of Voronoi tessellation where all nucleation sites are hexagonally close-packed and even spacing, $r_{hex}$. For general 2D cellular structures, a quantitative 'disorder parameter' δ can be defined using the minimum distance between centroid spacings '$s$' as (*20*):

$$\delta = \frac{s}{r_{hex}} \quad (1)$$

Examples of resulting Voronoi structures with δ from 0.1 to 1.0, and the corresponding cell size distributions are shown in **Fig. 1g**.

Remarkably, the biological materials, from trabecular bone to plant stems and fungi, generally fall into well-defined ranges of disorder (**Fig. 1i**), e.g.; woods and fungi from δ = 0.6 to 0.8; trabecular bone and dentin from δ = 0.55 to 0.65. Those biological materials associated with organisms constructing housing (corals, bee honeycomb) have relatively high order (δ = 0.9 to 0.97). We also find that common engineering metal and polymer foams, with structures dependent on surface tension effects (*21, 22*), are generally much more disordered (δ = 0.2 to 0.5). Regular hexagonal structural honeycombs (lightweight sandwich panels) have δ = 1.0. These results suggest there may be a certain *optimal* degree of disorder, which we refer to as 'pseudo-order', in biogenic cellular materials. The analysis in **Fig. 1i** is not intended to be exhaustive, but to highlight several representative materials and tissues from organisms spanning four Eukaryotic kingdoms.

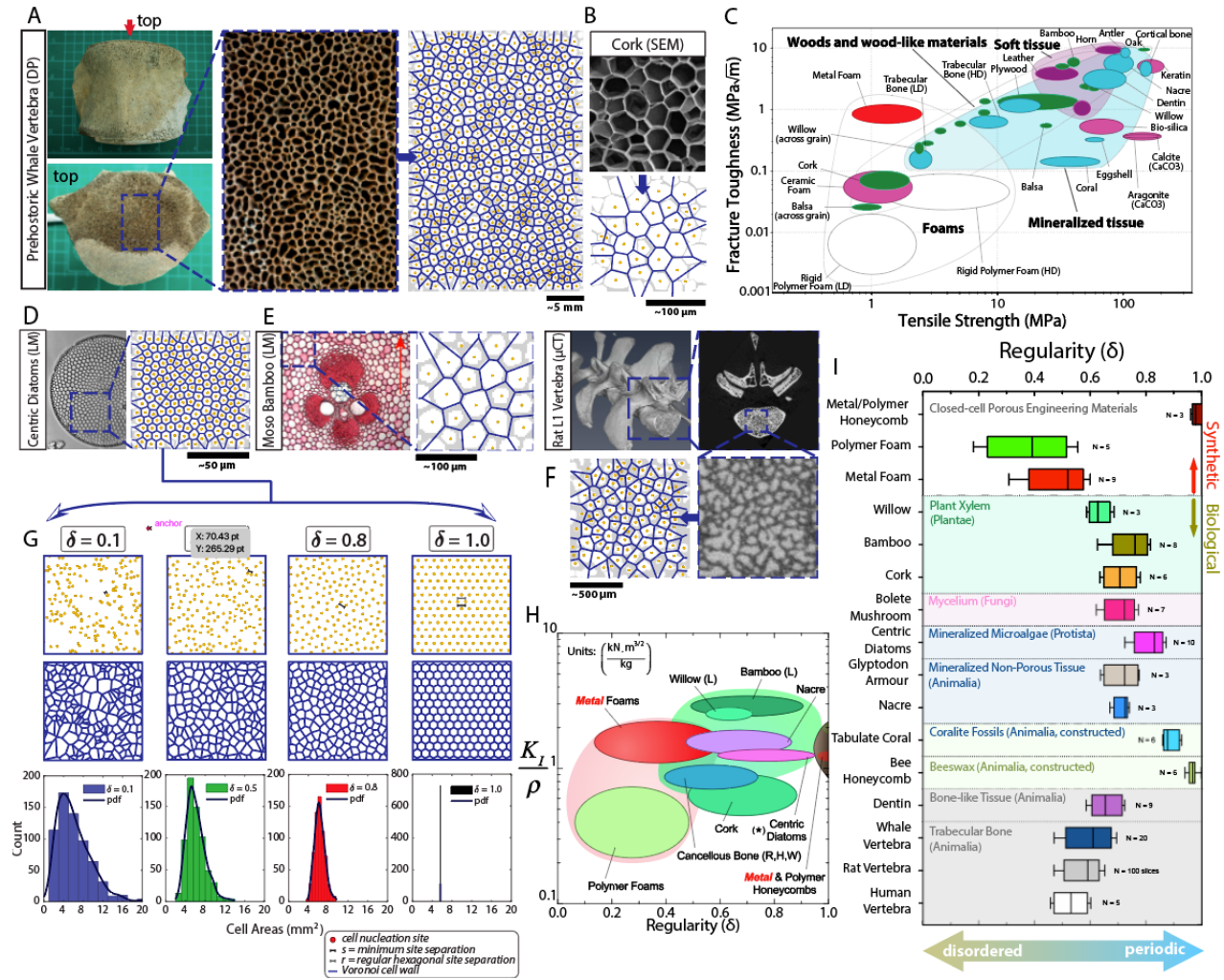

**Fig. 1**: Examples of biomaterials with disordered in-plane pore distributions and their superimposed Voronoi tessellations: (A) axial trabecular bone in whale vertebra; (B) radial cross-section of cork cell walls; (D) surface features in silica diatom frustule (coscinodiscus radiatus); (E) cross-section of parenchyma cells in Moso bamboo; (F) axial trabecular bone in rat vertebra. Images (B), (D) and (E) adapted from Refs. (*24-26*)), respectively. (G) Geometric effects of increasing regularity (δ) on Voronoi nucleation site distribution, cell morphology, and cell area distribution. (C) Material property map for fracture toughness and tensile strength in natural and foamed materials. Values sourced from CES EduPack (2013). (H) New material property chart for density-normalized fracture toughness demonstrating relationship with newly measured δ values. (I) δ values measured by image analysis for all surveyed cellular materials (sample details in SM). N describes number of images analyzed.

To compare the mechanical behaviour of these architectures, **Fig. 1c** presents the material property space for fracture toughness and tensile strength in biogenic and engineered cellular solids (*23*). The 'best' performance (high strength *and* damage tolerance) exists at the upper right-hand corner. It is clear that biogenic materials generally outperform engineered cellular solids, despite being composed of brittle constituents, such as calcium carbonate, apatites, silica, and cellulose. If we compare the normalized fracture toughness data from **Fig. 1c** for a range of common biological structural materials as a function of the disorder ($\delta$) (**Fig. 1h**), the highest specific toughness values are found for materials with the 'pseudo-ordered' $\delta$ values (0.6 to 0.8) found previously.

To date, there are no studies on the role that disorder plays in flaw tolerance or mechanical fracture of biogenic cellular materials, and few studies on the role of disorder in the mechanical properties of cellular materials generally. Zhu et al. and Sotomayor et al. used finite element (FE) modelling to simulate compressive elastic modulus in 2D honeycombs as a function of increasing disorder (*27, 28*). The primary failure mechanism of RHH in compression is either sequential plastic collapse of cell walls, or fracture and crushing of brittle cell walls, along close-packed directions, depending on the ductility of the parent material (*29, 30*). Studies show that point defects (filled cells or cut walls) can disrupt the sequential plastic collapse of cells, to increase damage delocalization in compression (*31-33*). FE models of impact compression have shown that increased cell irregularity in 2D honeycombs can increase the plateau stress, thereby improving the energy absorption capacity (*34-37*). There have been no studies that examine the role of disorder in tensile failure of cellular materials, though mode I tensile loading and fracture are common for biological structural materials.

In this work, we examine what role structural disorder contributes to toughening, particularly in tensile loading and fracture conditions, and whether the disorder (pseudo-order) of biogenic cellular materials may provide evolutionary advantage in tissue repair and organism survival. We have taken advantage of digital design and fabrication to generate 2.5D cellular models with systematically increasing degrees of disorder ($\delta$), for the same relative density and material composition.

**Results**

We have used high-resolution, multi-jet 3D printing to generate 2.5D honeycombs based on varying values of $\delta$. Modern digital fabrication tools provide the opportunity to systematically control disorder. Through uniaxial mechanical testing we compare the properties of disordered honeycombs (DH) with regularity from $\delta= 0.1$ to $0.9$, with the regular hexagonal honeycomb (RHH), $\delta= 1.0$. All throughout, relative density ($\bar{\rho} = 25\%$) and the number of nucleation sites (cells) are kept constant ($n = 314$). A CAD algorithm was built in Grasshopper (Rhinoceros 3D, USA) to convert base coordinates for each tessellation into a 3D solid (**Fig. 2a-d**), similar to the parametric approach of Frølich et al. for non-periodic tessellations in nacre (*38*). A series of finite element (FE) simulations was also employed to compare and expand on the mechanical performance of the disordered honeycombs. Further details on Voronoi regularity parameter definition and the FE modelling approach used to define fracture toughness are found in SM. Representative tensile stress-strain results are shown in **Fig. 2e**. The peak strength of RHHs ($\delta=1.0$) is higher than all DHs tested, yet also fracture in a single catastrophic event. In contrast,

DHs with 0.1 ≤ δ ≤ 0.8 fail in multiple steps by successive, non-catastrophic fracture with significant variations in crack direction. Each change of direction in the DH's path corresponds to a crack arresting event, visible as small dips in the stress-strain curve (**Movie M1** in SM), with the integrated area underneath the curve (shaded) indicating the total strain energy absorbed. While disordered honeycombs exhibit fracture events before regular honeycombs, their path to final fracture is prolonged, often producing higher energy absorption measurements than regular structures.

The relative effect of disorder on standard tensile properties can be seen by normalizing the DH values to those of the ordered RHH control (**Fig. 3**). The elastic modulus, E, remains largely unaffected by the changes in δ with the mean staying within 5% of the RHH value (**Fig. 3b**). The normalized tensile strength of the RHHs, however, are over 30% stronger than the most disordered δ=0.1 DHs. Interestingly, there is no statistical difference in $\sigma_{max}$ between the RHH and the δ=0.8; δ=0.7; and δ=0.6 DHs. This may be related to the fact that a near-uniform cell size is approached as δ → 1.0 (**Fig. S12, S13**) (*32*). The effect of disorder is even larger for the energy absorption capacity, with the similar effect that $U_v$ values for δ=0.8; δ=0.7; δ=0.6 are comparable to RHH values. These results show the 'pseudo-ordered' honeycombs of disorder 0.6 ≤ δ ≤ 0.8 exhibit comparable strength and energy absorption values as the regular honeycombs, but also multi-stage failure.

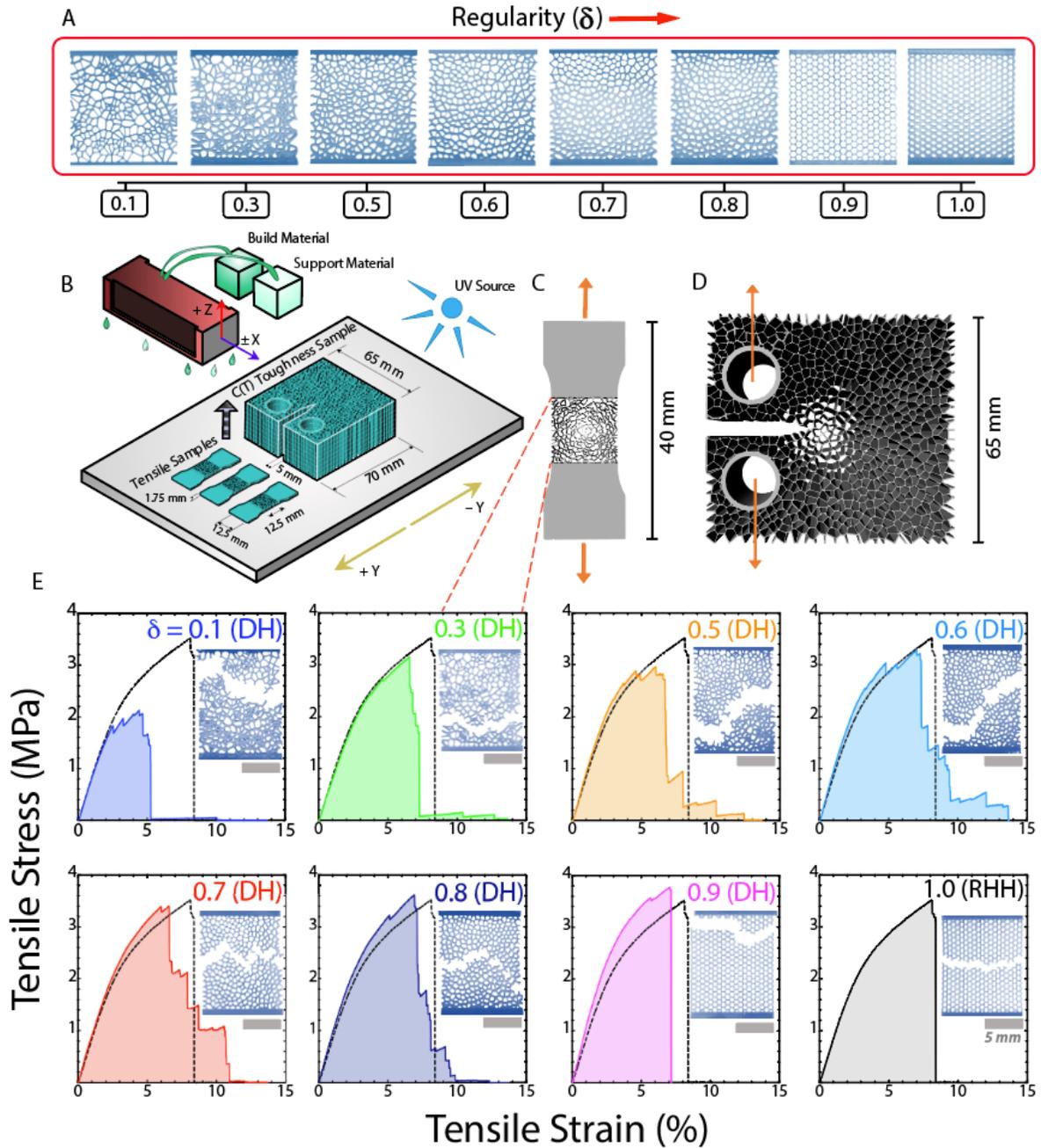

**Fig. 2:** (A) Photographs of printed dogbone gauge sections in tensile test coupons; (B) Schematic of printing process for (C) dogbone and (D) compact tension geometries. (E) Tensile stress-strain curves showing failure sequence in representative DH and RHH samples with insets showing photographs of the gauge section after fracture. Dotted lines show direct comparisons with the RHH.

Catastrophic fracture features an irrecoverable drop in load bearing capacity. The standard tensile properties in **Fig. 3b-d** are insufficient for describing the effect of disorder on fracture 'survivability' as a measure of damage tolerance, or non-catastrophic failure. Hence, three new 'survivability' metrics are defined in **Fig. 3a**: tensile strength, instantaneous strain, and energy absorption are measured at the first fracture event (subscript "1") and compared with the point of catastrophic failure (subscript "2", or "max" in the case of peak strength) in **Fig. 3e-g**. Their difference is plotted in **Fig. 3h-j**. We observe increased damage survivability in all properties relative to the RHH for $0.5 \leq \delta \leq 0.8$ DHs, with a remarkable 84% higher mean change in absorbed energy for $\delta=0.8$ (**Fig. 3j**). In other words, pseudo-ordered architectures not only match RHH strength, but they exceed its post-yield damage tolerance (i.e. survivability).

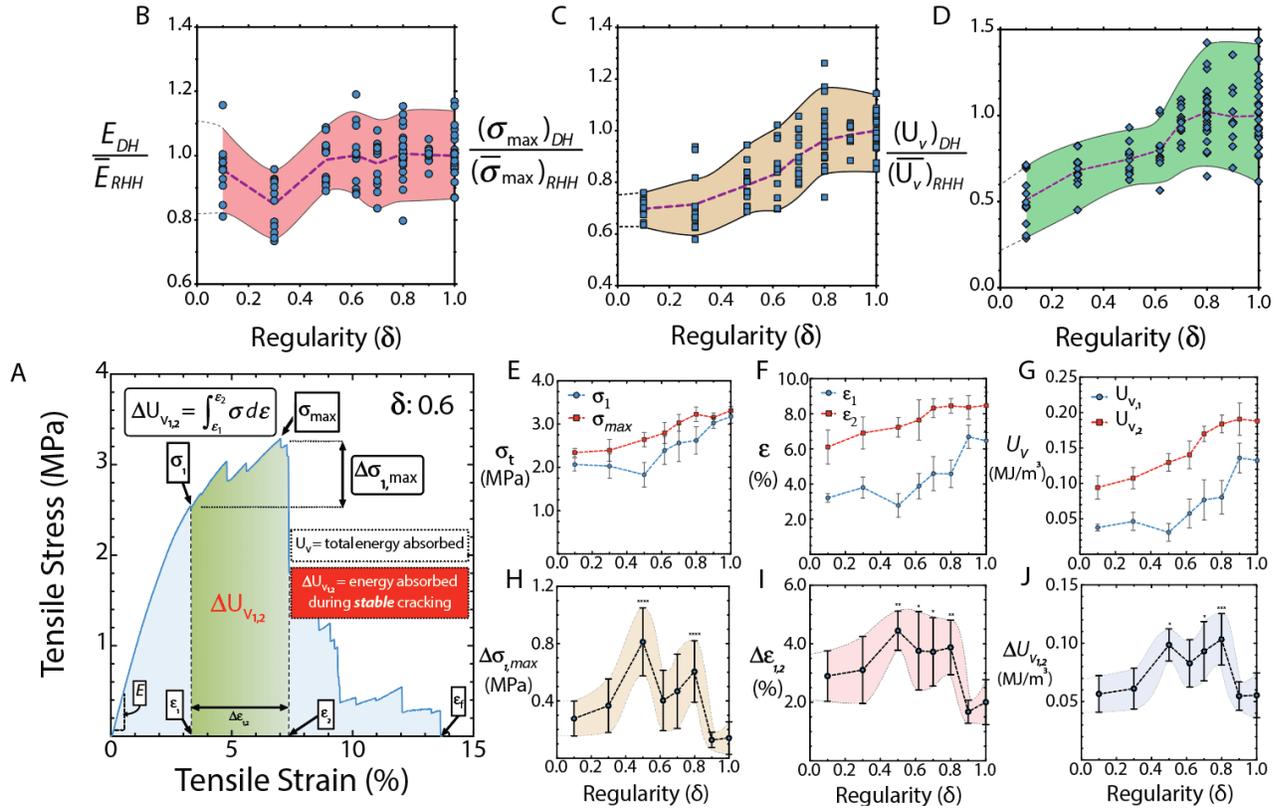

**Fig. 3:** Summary of uniaxial tensile properties in printed dogbone coupons. (A) Representative stress-strain response for $\delta = 0.6$ illustrates critical measurement locations (subscripts 1, 2 and max) for relevant properties. Change in energy absorption, $U_v$, during strain-hardening is shown as a shaded green area under the curve. Evolution with $\delta$ of normalized (B) Young's modulus; (C) tensile strength; and (D) energy absorption, respectively, plotted from individual tests (N=12 samples per $\delta$ value, with N = 24 for $\delta = 0.8$ and $\delta = 1.0$). Data is normalized against respective mean value from RHH tests. Growth in strength, ductility and energy absorption at critical points is shown in (E), (F), (G), respectively. Divergence between blue and red lines reveals strain-hardening survivability metrics depicted by differentials ($\delta$) in (H), (I), and (J). Error bars indicate standard deviation, while shaded regions are intended to guide the eye.

To gain further insight into the crack propagation mechanics we look to a more robust framework to quantify damage tolerance (*15, 39-44*). We employ elastic-plastic fracture mechanics (EPFM) on the larger uniaxial coupons outlined in **Fig. 2c** to capture incremental toughness changes during failure via the *J*-integral compliance method. By tracking tensile force and crack opening displacement (COD), crack paths can be calculated and compared to video-tracked crack lengths (**Fig. S16**). Here, $J$ is the energy release rate as the work done per unit area of crack advance (*45*).

The stress-intensity based evolution of fracture toughness, $K_J$, is plotted against relative crack extension in **Fig. 4a**. Derived from the $J$-integral, resistance curves of representative honeycombs are shown in the figure for each δ. Details of this derivation and fitting are presented in the SM sections A.5.2 and A.6. The quantity $\Delta a/b_0$ reflects the relative crack growth as it moves through the uncracked portion of material (initial dimension $b_0$). As such, the test was uniformized with data only considered up to 50% crack propagation (i.e. when $\Delta a/b_0 = 0.5$).

The fracture toughness is clearly influenced by the degree of disorder. In all tests, the honeycomb materials show toughening with crack extension, visualized by the modestly rising resistance curves in **Fig. 4a** (non-zero slope). We observe that DHs with δ > 0.3 have a higher slope than the reference RHH in the initial crack initiation region (up to $\Delta a/b_0 \approx 0.05$), whereas the RHH has a higher slope later in the curve. This is supported by the RHH fracture peak behaviour seen in the force-displacement curve of **Fig. S14,** where load drops were highly regular but of low magnitude and thus corresponded to individual ligament rupture. The fracture toughness values of pseudo-ordered DHs (δ ≥ 0.6) and RHHs converge when $\Delta a/b_0 > 0.3$, indicating that disorder has a diminished effect on compliance after some critical crack size. The most significant difference in fracture resistance is thus observed in the crack initiation stage.

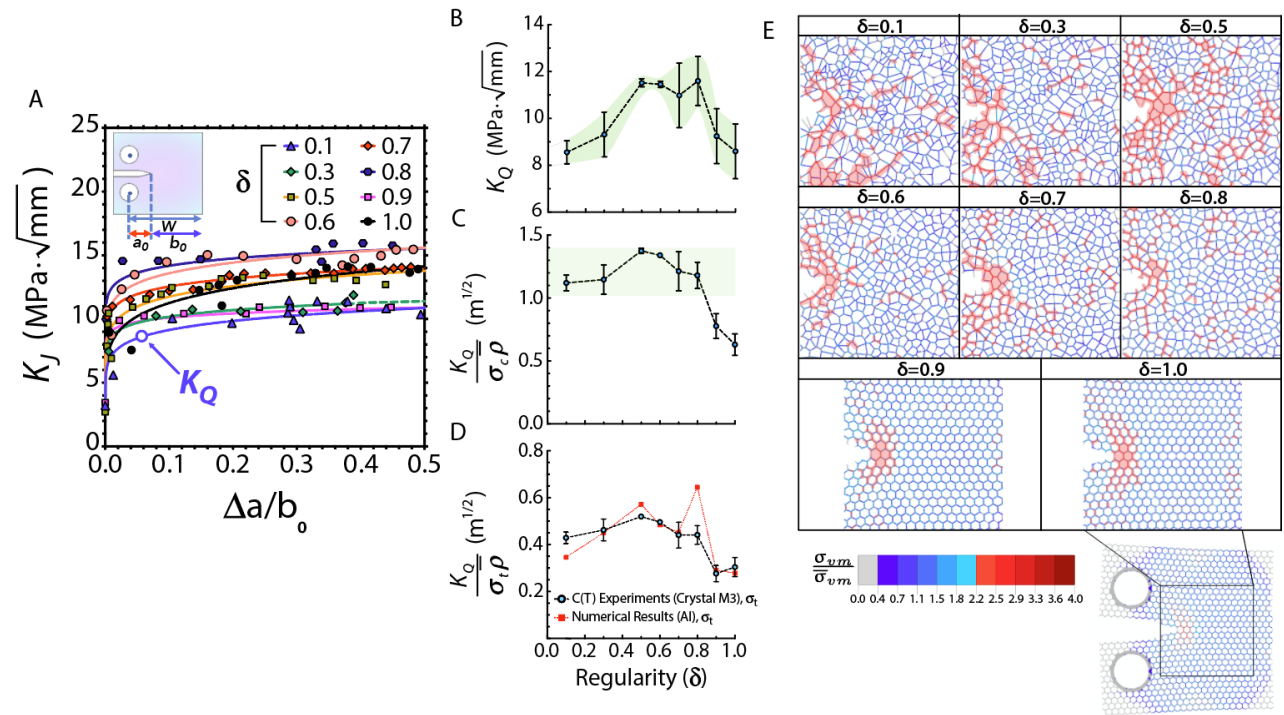

**Fig. 4:** Experimental and numerical C(T) results showing (A) the response of fracture toughness, $K_J$, against crack growth Δa, normalized by the initial un-notched sample width, $b_0$. Lines through data follow a conventional power law fit. (B) The effects of honeycomb disorder on mode I crack initiation toughness $K_Q$. Error bars represent standard deviation. Normalization yields (C) the compressive toughness efficiency, and (D) a comparison of numerical and experimental tensile fracture toughness

efficiency. (E) Normalized von mises stress contour maps from FEA taken in the timestep immediately before the first ligament fracture. Legend is normalized by the mean von mises stress summed from all beams in the sample. High-damage zones (continuously high $\sigma_{vm}$) are shaded red for visibility. FE simulations are all based on 2D 4-node CPE4R elements.

As with the tensile $U_v$ before, δ is shown to have a significant effect on honeycomb crack initiation toughness (mode I). This is denoted by parameter $K_Q$, plotted in **Fig. 4b** in lieu of the $K_{Ic}$ seen in bulk solids (See SM section A.6 for derivation). $K_Q$ values of pseudo-hexagonal DHs in the range 0.5 ≤ δ ≤ 0.8 consistently surpass that of regular honeycombs by 28-36%. Mean values of $K_Q$ are reported in Table S4, showing that δ = 0.8 reaches a fracture toughness of 0.37 MPa.$m^{1/2}$. Further normalization by mean compressive strength and relative density yields a mode I "toughness efficiency", $K_Q/(\sigma_c \bar{\rho})$. **Fig. 4c** demonstrates how the efficiency parameter for all DHs is superior to that of the RHH.

Fracture toughness findings were cross-examined by a newly developed finite element (FE) model (method and rationale in SM). **Figs. 4d,e** display numerical results for analogous tests on 2D aluminum honeycombs with identical geometry to the experimental coupons. Remarkably, the toughness efficiency with respect to *tensile* strength ($K_Q/(\sigma_t \bar{\rho})$) exhibits an overlapping trend over experimental data (**Fig. 4d**). **Fig. 4e** compares a side-view of the RHH and DH samples from numerical C(T) tests in the timestep immediately ahead of the first fracture event. It is clear in the superimposed von Mises stress maps that the RHH concentrates stress in the row of cells directly ahead of the crack tip. DHs instead show a much wider dispersion of high stress zones (shaded red for clarity) across the entire sample area, which aligns with previous observations of strain delocalization (*28, 46*). All together, these behaviours suggest that structural disorder plays an important role in toughening of both brittle and ductile material systems alike.

**Discussion**

This work demonstrates for the first time, through experiment and simulation, that spatial disorder in cellular solids itself can have dramatic effects on fracture behaviour compared to ordered, regular hexagonal honeycombs (RHH). Disorder helps to delocalize the stress at cracks, inhibits catastrophic crack propagation through path deflection (and multiple cracks), and thus increases effective fracture toughness. As shown experimentally (**Fig. 2e**) and numerically (**Fig. S21**), the fracture paths in the RHHs follow the predominantly straight trajectories of catastrophic fracture. In RHHs, the stress field in any cell ahead of an advancing crack tip is identical. Similarly, in compression, RHHs are known to fail through a propagating cascade of collapse (*31*). In contrast, the DH cells cause the crack tip path to deflect under influence of unique asymmetric stress fields that emerge after each ligament fracture (**Fig. 4g**). As such, each tip redirection is a pause in crack propagation, promoting DH stability. The most significant path deflections were universally seen in the 0.6 ≤ δ ≤ 0.8 range, while straight-line "fast fracture" sections of crack growth only occur when δ < 0.5 and δ ≥ 0.9 (**Fig. 2e**).

The measured toughness and tensile strength values demonstrate how disordered tessellations in the range of 0.6 ≤ δ ≤ 0.8 can match the strength of ordered RHHs while increasing the work of fracture, and 'survivability' through crack deflection. The disorder range around δ=0.8 in particular shows a mode I fracture toughness $K_Q$ that is 36% higher than δ=1.0 (for RHHs). For the first time, these results show that for all other parameters equal (average cell size, area fraction and relative density) there are real mechanical advantages to disorder.

Time-correlated video stills (**Fig. S17**) reveal how toughness enhancements appear to stem from crack tip blunting and deflection mechanisms analogous to those found in biological composites like wood, nacre and bone (*10, 11, 16, 17*). We adopt Kruzic et al.'s formalism for intrinsic and extrinsic toughening mechanisms (*16*). Plastic wall bending and "crack diffusion" (sudden fracture bursts) (**Fig. S17b,c**) are deemed as intrinsic events, affecting crack motion ahead of the tip to resist further damage. Small, progressive fracture bursts allow the structure to retain load carrying capacity during failure. Secondly, crack tip bridging is an extrinsic mechanism. Ligament bridges help to further stabilize crack growth in the bulk by effectively holding together splitting regions of the honeycomb. In other words, for δ ≤ 0.8 there is a flaw tolerance, in the form of 'reserve' structural loading capacity despite the onset of damage. Indeed, Pro and Barthelat recently summarized the universal importance of weak interfaces in the toughness of biological materials (*15*). We suggest that 'weak', sacrificial cell walls in our DHs provide energetically preferred cracking planes, to lead the crack tip away from a single catastrophic path.

Toughening mechanisms and flaw tolerance in biological cellular materials, such as trabecular bone and dentin, have important consequences for tissue repair, and survival. After trauma, it is critical for animals to avoid catastrophic bone fracture for any chance to survive. **Fig. 5** depicts hypothetical stages of fracture in the femoral trochanter of mammalian long bone, together with a typical tensile stress-strain curve for δ= 0.8 DH. Fracture events are coupled to a series of stress drops. As such, fracture #1 shows a near-imperceptible loss of loading capacity and quick recovery of plasticity. This could relate to internal micro-cracks within trabecular bone (panel 1, and Fazzalari et al. (*47*)). Panels 2 – 4 depict partial "greenstick" fractures, whereas 5 shows a severed bone. The containment of fracture partway through the bone cross-section may be seen as an evolutionary design goal, prolonging an injured animal's ability to bear weight on the injury. Such flaw size control has been attributed to crack bridging mechanisms (*10, 15, 17, 48, 49*), and, recently, trabecular surface plasticity (*19*).Toughening mechanisms for crack deflection and damage delocalization result in more, smaller cracks, instead of a single, catastrophic crack and large crack opening displacement (COD).

Tissue repair mechanisms are scale dependent and have an *absolute* size limit for defects to be repaired. For example, in bone the 'primary healing' mechanism is through the Basic Multicellular Units (BMUs) of osteoclasts and osteoblasts (*10, 18, 19, 49*) to tunnel through and repair damage by bone degradation and re-deposition. The BMUs can bridge cracks, but only for COD of less than 100-200 μm (**Fig. 5c**) (*49*). Micro-damage in bone (COD < 100 μm) is common, easily repaired, and part of normal bone remodelling. However, a large COD requires the secondary healing mechanism, requiring much more time (and callus formation). The damage delocalization in disordered honeycombs limits absolute crack size, and thereby also keeps the associated COD gap below a threshold size in bone to enable the primary repair mechanism, and better chance of survival.

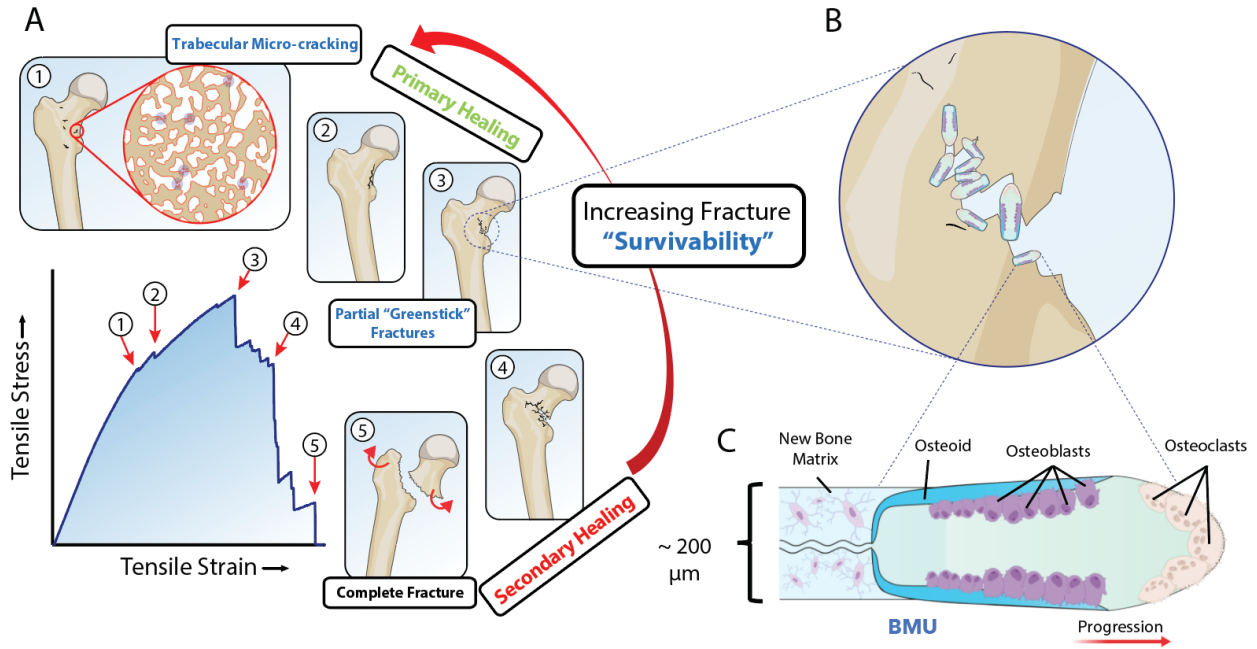

**Fig. 5:** A) Schematic illustrations of fracture events with increasing severity in a femoral bone, compared to fractures in the stress-strain curve of a δ=0.6 DH. Fractures are categorized by 'ease' of repairability by (B) basic multicellular units (BMUs) that can traverse cracks in bone up to a limited scale (100-200 μm), as a mechanism for remodeling and defect repair. (C) Detail illustration depicts the finite structure and scale of the BMU. Distributed, delocalized damage enables effective tissue repair, whereas single, large cracks (large COD) require much more time and biological resources (secondary repair mechanism), and less survivable.

Our study is the first to quantify the disorder in biological cellular materials. We hypothesize that the distinct ranges of pseudo-order we have identified in biological cellular materials for a wide range in organisms is not accidental as it provides mechanical advantage in fracture toughness. In addition to the evolutionary advantage in limiting catastrophic fracture in trauma, we suggest that damage delocalization can enable tissue repair and thereby improve chances of survival. Since we find this similar degree of structural disorder across organisms from the Eukaryotic kingdoms of Animalia, Plantae, Fungi, in our analysis of plant tissue, mammalian bone, sponges and fungi (polypores) (**Fig. 1h,i**), we suggest it is an example of convergent evolution. *In other words, tailored disorder offers an inherent mechanical advantage that is common to all structural cellular solids, and therefore it appears broadly throughout nature.*

Our results suggest that structural cellular materials should not be too highly ordered for optimal mechanical performance. However, there are examples of well-ordered cellular materials in nature. The bee honeycomb and coral examples in **Fig. 1i** are both relatively well-ordered with cells of uniform size. We can speculate that their chief function is to house insect pupa, and polyps, respectively, over mechanical performance. Also, there are many examples of biological cellular materials for structural color, such as in wings of many butterfly species, which require well-defined periodicity for Bragg diffraction. Despite this, certain benefits of disorder in biogenic structural colour materials have also been found. Zollfrank et al. define a "tailored disorder" for optical performance (*50*), in particular that disordered structures of materials with deficient refractive index – like chitin – achieves advanced photonic properties. This is shown in the bright isotropic structural colour of the ultra-white beetle which features disordered, pseudo-photonic scales (*51*).

It seems remarkable that disorder in engineered cellular materials, such as aluminum or polymer foams, has generally been overlooked in structural design. Our work shows the disorder in commercial foams is often much higher than found in biogenic materials. As a result, we suggest that 'tailored' structural disorder should be considered as a new design parameter for cellular, architected materials, independent of cell size, connectivity and relative density. Additive manufacturing and digital design techniques for structural optimization (such as topology optimization) allow exploration of a vast new parameter and performance space for architected materials. Recently, engineered hierarchical materials have incorporated toughening mechanisms at the microscale (*52-55*). We suggest that, as for biological cellular materials, composite design can incorporate 'designed disorder' in combination with other toughening mechanisms, such as weak interfaces, crack bridging and gradients.

**Conclusions**

We have quantified structural disorder in biological cellular materials, and identified limited ranges of disorder which are distinct from those found in typical engineered materials. Spatial disorder in 2D honeycomb significantly increases fracture toughness and flaw tolerance, to limit catastrophic fracture. Disordered honeycombs in the range of $0.6 \leq \delta \leq 0.8$, observed in our biological examples, show elevated fracture toughness and similar strength and stiffness compared to ordered hexagonal controls. The introduction of tailored disorder ($\delta=0.8$) to a 2D honeycomb of identical relative density enables a capacity for strain hardening, damage delocalization and limiting catastrophic failure. Through increased fracture toughness and damage delocalization, we suggest that the pseudo-order of biological cellular materials likely offer evolutionary advantage in survivability, both in limiting fracture and enabling tissue repair. In turn, designed disorder is identified as a potent design tool for future architected materials in engineering applications.

**Acknowledgments:** We thank Zachary Fishman and Prof. Cari Whyne of the Sunnybrook Research Institute for their expertise and provision of experimental x-ray µCT data for vertebral rat specimens. We are indebted to Dr. Dan Grozea for his mechanical testing expertise and gratefully acknowledge Prof. Marc Grynpas for his useful discussions and insights on trabecular bone mechanics.

**Funding:** This work was supported through the Natural Sciences and Engineering Research Council of Canada (NSERC) Discovery Program #06760 (BDH). D.A.v.E. was supported by the Queen Elizabeth II Graduate Scholarship in Science and Technology.

**Competing interests:** The authors declare no competing interests.

**Data and materials availability:** All data is available in the main text or the supplementary materials